\documentclass[aps,prl,groupedaddress,twocolumn,floatfix,showkeys,showpacs]{revtex4}
\usepackage{graphicx}

\def\be{\begin{equation}}
\def\ee{\end{equation}}
\def\ba{\begin{eqnarray}}
\def\ea{\end{eqnarray}}
\def\bc{\begin{center}}
\def\ec{\end{center}}

\begin{document}

\title{Propagation of edge magnetoplasmons in semiconductor quantum-well structures}

\author{S. A. Mikhailov}

\affiliation{Mid-Sweden University, ITM, Electronics Design Division, 851 70 Sundsvall, Sweden}

\date{\today}

\begin{abstract}
The wavelength and the propagation length of the edge magnetoplasmons, running along the edge of a two-dimensional electron layer in a semiconductor quantum-well structure are theoretically studied as a function of frequency, magnetic field, electron density, mobility, and geometry of the structure. The results are intended to be used for analysis and optimization of operation of recently invented quantum-well microwave spectrometers operating at liquid-nitrogen temperatures (I. V. Kukushkin {\em et al}, Appl. Phys. Lett. {\bf 86}, 044101 (2005)).
\end{abstract}

\pacs{73.20.Mf, 78.67.-n, 84.40.-x}
\keywords{edge magnetoplasmons, microwave photoresponse, detection and spectroscopy}

\maketitle

Recently, a new method of detection and spectroscopy of microwave radiation has been discovered \cite{Kukushkin04a,Mikhailov-SPIE04,Kukushkin05a}. The method is based on excitation and interference of a special type of plasma waves -- edge magnetoplasmons (EMP) \cite{Mast85,Glattli85,Volkov85b,Volkov86b,Fetter85,Fetter86a,Volkov88,Allen83,Wassermeier90,Grodnensky90,Volkov91,Peters91,Ashoori92,Mikhailov92,Talyanskii93,Zhitenev93,Aleiner94,Tonouchi94,Dahl95,Mikhailov95a,Balaban98,Mikhailov00a} -- in a semiconductor quantum-well structure. The EMPs propagate in a two-dimensional (2D) electron system along its edge and possess two important physical properties. First, their frequency is tunable across a very broad frequency range (experimental observation of EMPs have been reported at frequencies from $\sim 1$ kHz \cite{Peters91} up to $\sim 1$ THz \cite{Allen83}). Second, in contrast to the bulk plasmons and magnetoplasmons, their damping is small not only at $\omega\tau\gg 1$ but at any value of $\omega\tau$, if the applied magnetic field is sufficiently strong ($\omega_c\tau\gg 1$). Here $\omega$ and $\omega_c$ are the microwave and the cyclotron frequencies, and $\tau$ is the momentum relaxation time. These two EMP properties open unique opportunities to build resonant (frequency sensitive) detectors and spectrometers for the microwave and terahertz frequency range \cite{Kukushkin04a,Mikhailov-SPIE04,Kukushkin05a,Dorozhkin05}. 

In the experiments \cite{Kukushkin04a,Mikhailov-SPIE04,Kukushkin05a}, a quantum-well semiconductor sample with a 2D electron gas (EG) is placed in an external magnetic field and irradiated by microwaves. Radiation excites EMPs in the near-contact regions of the sample. EMPs, emitted from different contacts, propagate along the edge of the 2DEG and interfere with each other. This leads to the microwave induced photovoltage between pairs of contacts, proportional to $|1+\exp(iqL)|^2$, where $L$ is the distance between the contacts and $q(\omega)$ is the EMP wavevector directed along the edge of the 2DEG. This photovoltage oscillates as a function of the applied magnetic field, with the oscillation amplitude and period proportional to the microwave power and wavelength respectively, which allows one to use the effect for frequency-sensitive detection and spectroscopy of radiation.

Due to scattering of electrons in the sample, the wavevector $q(\omega)$ is a complex function, with the real part $q'(\omega)$ determining the EMP wavelength and the imaginary part $q''(\omega)$ -- their decay length. For proper device operation, $q'$ must be larger than $q''$ and the parameter $e^{-q''L}$ should not be too small. The quantities $q'(\omega)$ and $q''(\omega)$, apart from the radiation frequency, also depend on the magnetic field $B$, the 2D electron density $n_s$, the mobility of 2D electrons $\mu=e\tau/m^\star$ (and hence on the temperature) and dielectric environment of the sample. The goal of this letter is to quantitatively analyze these dependencies, so that this information could be used for analysis and optimization of device operation.

\begin{figure}
\includegraphics[width=8.4cm]{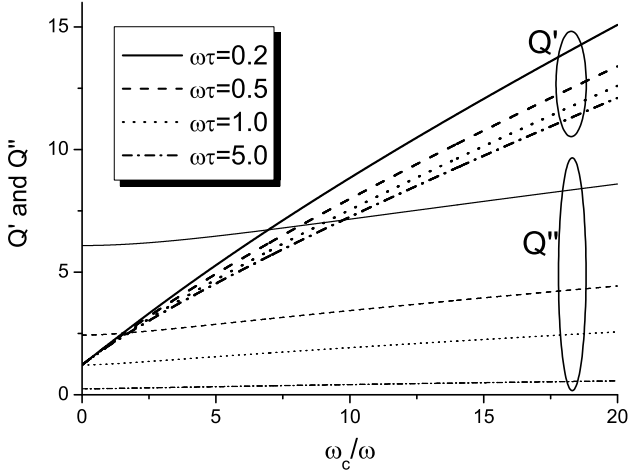}%
\caption{\label{qvswc}
Real and imaginary parts of the dimensionless EMP wavevector $Q$, defined by Eq. (\ref{Q}), as a function of $\omega_c/\omega$ at different values of $\omega\tau$. }
\end{figure}

First, consider a 2DEG lying on the boundary of a semiconductor substrate (the dielectric constant $\kappa$) with air. In this case, the EMP dispersion equation \cite{Volkov88} can be written as $Y=G(X)$, where $Y=(\omega+i/\tau)/\omega_c$, 
\be
G(X)=\tanh\left\{\frac 1\pi\int_0^{\pi/2}dt\ln\left(1+\frac {X}{\sin t}\right)\right\},
\ee
\be
X=\frac {2\pi n_se^2q(\omega+i/\tau)}{m^\star\bar\kappa\omega(\omega_c^2-(\omega+i/\tau)^2)},
\ee
and $\bar\kappa=(\kappa+1)/2$ [the bulk magnetoplasmon (BMP) dispersion relation \cite{Chiu74} has the form $1+X=0$]. If $F=G^{-1}$ is the inverse function of $G$, $X=F(Y)$, the above equations can be written as
\be
Q\equiv\frac {2\pi n_se^2q}{m^\star\bar\kappa\omega^2}=
\frac{\omega_c^2-(\omega+i/\tau)^2}{\omega(\omega+i/\tau)}
F\left(\frac{\omega+i/\tau}{\omega_c}\right).
\label{Q}
\ee
One sees that both real and imaginary parts of the wavevector $q$ are inversely proportional to the electron density, and that the dimensionless complex wavevector $Q$ is a function of two parameters, $\omega_c\tau$ and $\omega\tau$. Figure \ref{qvswc} shows $Q$ as a function of dimensionless magnetic field $\omega_c/\omega$ at several different values of $\omega\tau$. The imaginary part $Q''$ can be well approximated, for all curves in Figure \ref{qvswc}, by the formula
\be
Q''\approx \frac{1.217}{\omega\tau}\left(1+0.09\frac{\omega_c\tau}{\sqrt{1+\omega^2\tau^2}}\right).
\ee

\begin{figure}
\includegraphics[width=8.4cm]{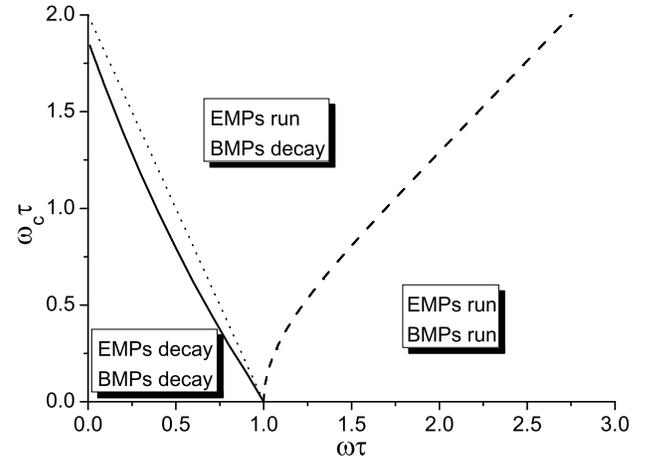}%
\caption{\label{empbmp}
Solid and dashed curves separate the regions of existence $(q'>q'')$ and decaying $(q'<q'')$ of the edge and bulk magnetoplasmons (BMP), respectively. The dotted line is given by Eq. (\ref{bound}).}
\end{figure}

As seen from Figure \ref{qvswc}, if $\omega\tau>1$, $Q'$ is always bigger than $Q''$, EMPs can propagate in the sample, and their interference can be observed. If $\omega\tau<1$, $Q'$ exceeds $Q''$ if the value of $\omega_c\tau$ is sufficiently large \cite{Volkov88}. Figure \ref{empbmp} quantitatively illustrates this EMP property: the solid curve separates the regions where EMPs can propagate in the sample $(q'>q'')$ and where they decay $(q'<q'')$. Roughly, the EMP observability condition can be written as 
\be
(\omega+\omega_c/2)\tau>1
\label{bound}
\ee
(the dotted line in Figure \ref{empbmp}): if $\omega\tau$ is smaller than 1, this can be compensated by applying a finite magnetic field. For example, in a GaAs/AlGaAs heterostructure at liquid nitrogen temperatures ($\mu\simeq 10^5$ cm$^2$/Vs) and at $f\simeq 100$ GHz the $\omega\tau$ parameter exceeds unity already at zero $B$, $\omega\tau=2.4$. At room temperatures ($\mu\simeq 10^4$ cm$^2$/Vs) and $f\simeq 100$ GHz, $\omega\tau=0.24$ and one needs to apply about 1.4 T to revive EMPs. At higher frequencies the corresponding required magnetic fields are lower ($\simeq 1$ T at 200 GHz and $\simeq 70$ mT at 400 GHz). 

Writing the BMP dispersion relation $1+X=0$ in the form $Q\equiv Q'+Q''=F_b(\omega\tau, \omega_c\tau)$ and equating $Q'$ and $Q''$, we can also get a curve on the plane $\omega_c\tau$ vs $\omega\tau$, separating the regions of propagating and damping the bulk plasma waves. It is shown for comparison by the dashed curve in Figure \ref{empbmp}.

So far, we have analyzed the simple case of the 2DEG lying on the surface of a semi-infinite semiconductor substrate. Now consider a more realistic model air -- semiconductor overlayer (dielectric constant $\kappa$, thickness $d_1$) -- 2DEG -- semiconductor substrate (dielectric constant $\kappa$, thickness $d_2$) -- air. In this case it is convenient to measure the wavevector in units of $d^{-1}$ and the frequencies in units of 
\be
\omega_0=\sqrt{\frac{2\pi n_se^2}{m^\star d}},
\ee
where $d=d_1+d_2$ is the full thickness of the semiconductor layer (typically $\simeq 0.4$ mm). Then the EMP dispersion equation \cite{Volkov88} can be written in the form
\be
qd={\cal F}\left(\omega/\omega_0,\omega_c/\omega_0,\omega_0\tau,\kappa, d_1/d\right),
\label{qd}
\ee
where ${\cal F}$ is a complex function of the listed parameters. In a typical quantum-well structure the frequency $\omega_0$ lies in the GHz range,
\be
\frac{\omega_0}{2\pi}=24.5 \ {\rm GHz}\sqrt{\frac{n_s(10^{11}\ {\rm cm}^{-2})}{d({\rm mm})}};\label{es1}
\ee
here we have used the effective mass of GaAs electrons ($m^\star=0.067m_0$).

\begin{figure}
\includegraphics[width=8.4cm]{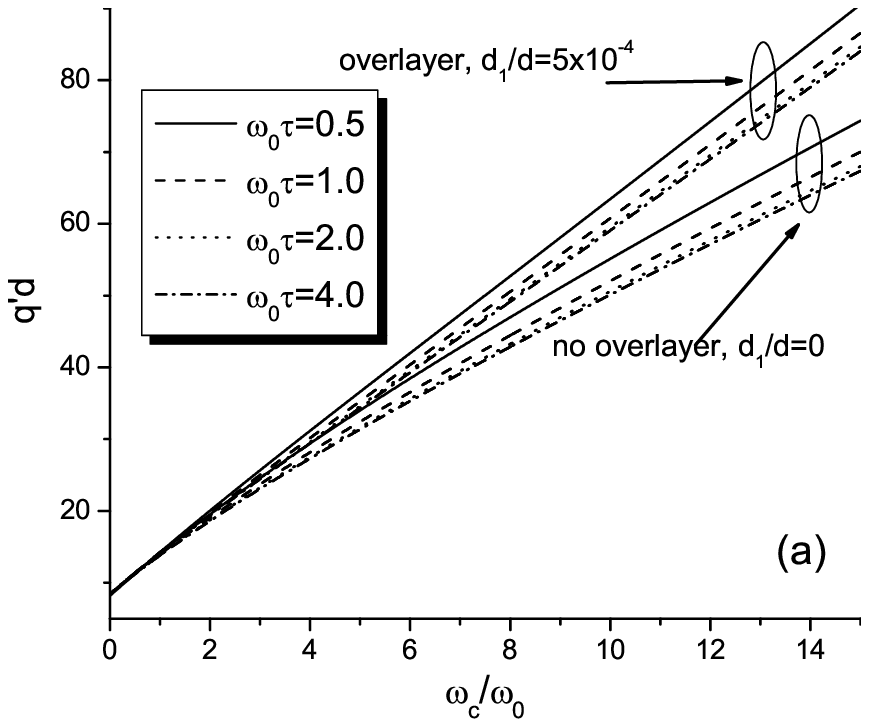}\\
\includegraphics[width=8.4cm]{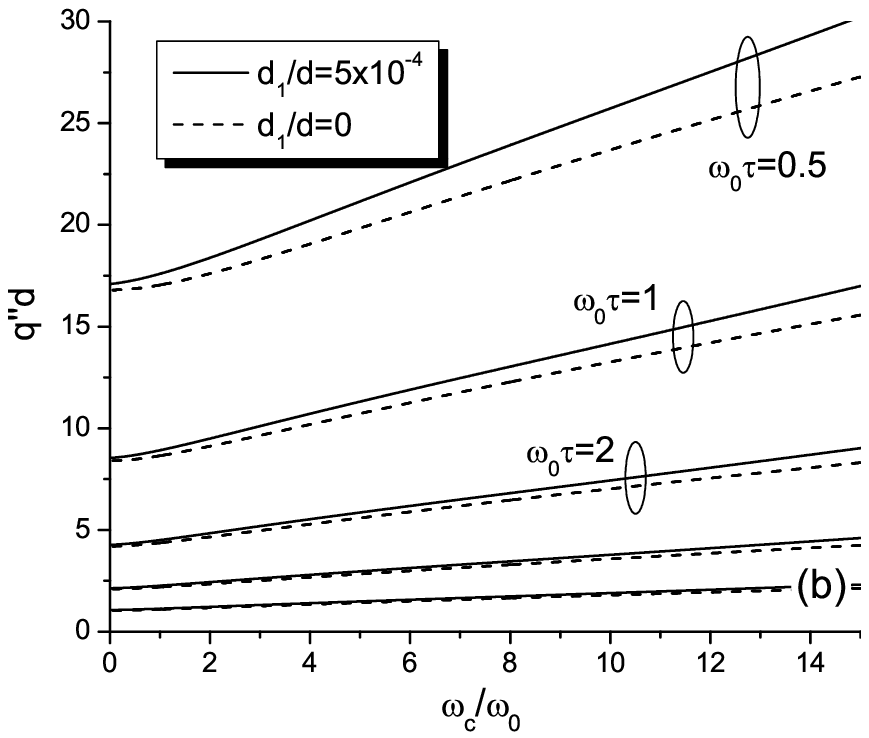}%
\caption{\label{qdW1} 
(a) Real and (b) imaginary parts of the dimensionless wavevector $qd$ as a function of dimensionless magnetic field $\omega_c/\omega_0$ at different values of $\omega_0\tau$ in structures with ($d_1/d=5\times 10^{-4}$) and without ($d_1/d=0$) the overlayer. Two lowest pairs of curves in Figure (b) correspond to $\omega_0\tau=4$ and $\omega_0\tau=8$. The microwave frequency is $\omega/\omega_0=1$, which corresponds to about $50$ GHz at experimental parameters \protect\cite{Kukushkin04a,Mikhailov-SPIE04,Kukushkin05a}.}
\end{figure}

\begin{figure}
\includegraphics[width=8.4cm]{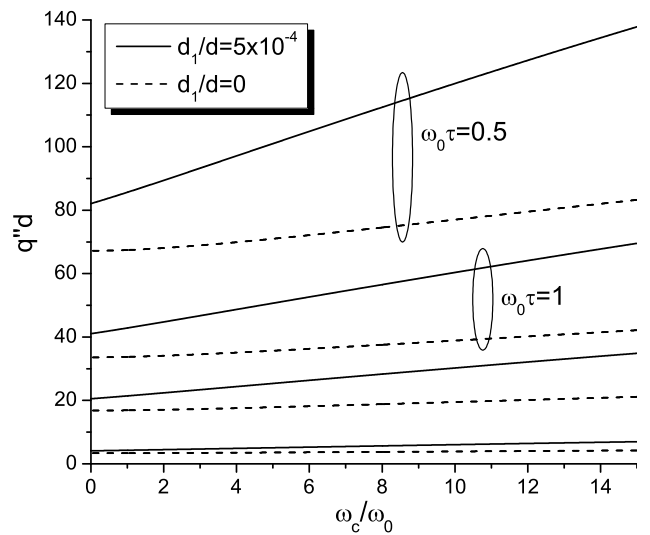}%
\caption{\label{qdW2} 
The same as in Figure \ref{qdW1}b but for $\omega/\omega_0=4$. The pairs of solid and dashed curves correspond to (from top to bottom): $\omega_0\tau=0.5$, 1, 2, and 10. }
\end{figure}

Analysing the EMP dispersion equation (\ref{qd}) we have found that, under experimental conditions \cite{Kukushkin04a,Mikhailov-SPIE04,Kukushkin05a}, the finite substrate thickness $\simeq 0.4$ mm practically does not influence the $q(\omega_c)$ dependencies. Results obtained for $d_2=0.4$ mm and $d_2=\infty$ are almost the same. In contrast, taking into account the finite overlayer thickness $d_1$ significantly changes the wavelength and especially the decay length of EMPs, see Figures \ref{qdW1} and \ref{qdW2} ($d_1$ is about 0.2 $\mu$m in a typical quantum-well structure and was usually considered to be negligibly small). The difference between the structures without ($d_1=0$) and with the overlayer ($d_1/d=5\times 10^{-4}$) increases with $B$ and $\omega$, and is larger in samples with lower $\omega_0\tau$. Physically, this is explained by a stronger localization of the EMP field near the edge of the 2DEG at higher frequencies and magnetic fields. As a result, the effective dielectric constant, which enters the formula for the wavevector $q$, see Eq. (\ref{Q}), varies from $\bar\kappa=(\kappa+1)/2$ to $\kappa$ with growing $B$ and/or $\omega$. 


To conclude, we have analyzed the real and imaginary parts of the EMP wavevector $q$ as a function of the microwave frequency, magnetic field, quality and dielectric environment of the samples. We have shown that the overlayer thickness significantly influences the wavevector, {\em reducing} the EMP propagation length and thus {\em suppressing} the output photovoltage signal. The finite overlayer thickness should therefore be taken into account for proper quantitative understanding of experimental data and for optimal design of detectors and spectrometers based on the considered effect.

\begin{acknowledgments}
The work was supported by the Swedish Research Council, the Swedish Foundation for International Cooperation in Research and Higher Education (STINT), and INTAS. The author thanks Igor Kukushkin, Jurgen Smet and Klaus von Klitzing for numerous discussions and information about running experiments.
\end{acknowledgments}


\begin{thebibliography}{27}
\expandafter\ifx\csname natexlab\endcsname\relax\def\natexlab#1{#1}\fi
\expandafter\ifx\csname bibnamefont\endcsname\relax
  \def\bibnamefont#1{#1}\fi
\expandafter\ifx\csname bibfnamefont\endcsname\relax
  \def\bibfnamefont#1{#1}\fi
\expandafter\ifx\csname citenamefont\endcsname\relax
  \def\citenamefont#1{#1}\fi
\expandafter\ifx\csname url\endcsname\relax
  \def\url#1{\texttt{#1}}\fi
\expandafter\ifx\csname urlprefix\endcsname\relax\def\urlprefix{URL }\fi
\providecommand{\bibinfo}[2]{#2}
\providecommand{\eprint}[2][]{\url{#2}}

\bibitem[{\citenamefont{Kukushkin et~al.}(2004)\citenamefont{Kukushkin, Akimov,
  Smet, Mikhailov, von Klitzing, Aleiner, and Falko}}]{Kukushkin04a}
\bibinfo{author}{\bibfnamefont{I.~V.} \bibnamefont{Kukushkin}},
  \bibinfo{author}{\bibfnamefont{M.~Y.} \bibnamefont{Akimov}},
  \bibinfo{author}{\bibfnamefont{J.~H.} \bibnamefont{Smet}},
  \bibinfo{author}{\bibfnamefont{S.~A.} \bibnamefont{Mikhailov}},
  \bibinfo{author}{\bibfnamefont{K.}~\bibnamefont{von Klitzing}},
  \bibinfo{author}{\bibfnamefont{I.~L.} \bibnamefont{Aleiner}},
  \bibnamefont{and} \bibinfo{author}{\bibfnamefont{V.~I.} \bibnamefont{Falko}},
  \bibinfo{journal}{Phys. Rev. Lett.} \textbf{\bibinfo{volume}{92}},
  \bibinfo{pages}{236803} (\bibinfo{year}{2004}).

\bibitem[{\citenamefont{Mikhailov et~al.}(2004)\citenamefont{Mikhailov,
  Kukushkin, Smet, and von Klitzing}}]{Mikhailov-SPIE04}
\bibinfo{author}{\bibfnamefont{S.}~\bibnamefont{Mikhailov}},
  \bibinfo{author}{\bibfnamefont{I.}~\bibnamefont{Kukushkin}},
  \bibinfo{author}{\bibfnamefont{J.}~\bibnamefont{Smet}}, \bibnamefont{and}
  \bibinfo{author}{\bibfnamefont{K.}~\bibnamefont{von Klitzing}}, in
  \emph{\bibinfo{booktitle}{Passive Millimetre-Wave and Terahertz Imaging and
  Technology}}, edited by
  \bibinfo{editor}{\bibfnamefont{R.}~\bibnamefont{Appleby}},
  \bibinfo{editor}{\bibfnamefont{J.~M.} \bibnamefont{Chamberlain}},
  \bibnamefont{and} \bibinfo{editor}{\bibfnamefont{K.~A.}
  \bibnamefont{Krapels}} (\bibinfo{publisher}{SPIE},
  \bibinfo{address}{Bellingham, WA}, \bibinfo{year}{2004}), vol.
  \bibinfo{volume}{5619} of \emph{\bibinfo{series}{Proceedings of SPIE}}, pp.
  \bibinfo{pages}{187--197}.

\bibitem[{\citenamefont{Kukushkin et~al.}(2005)\citenamefont{Kukushkin,
  Mikhailov, Smet, and von Klitzing}}]{Kukushkin05a}
\bibinfo{author}{\bibfnamefont{I.~V.} \bibnamefont{Kukushkin}},
  \bibinfo{author}{\bibfnamefont{S.~A.} \bibnamefont{Mikhailov}},
  \bibinfo{author}{\bibfnamefont{J.~H.} \bibnamefont{Smet}}, \bibnamefont{and}
  \bibinfo{author}{\bibfnamefont{K.}~\bibnamefont{von Klitzing}},
  \bibinfo{journal}{Appl. Phys. Lett.} \textbf{\bibinfo{volume}{86}},
  \bibinfo{pages}{044101} (\bibinfo{year}{2005}).

\bibitem[{\citenamefont{Mast et~al.}(1985)\citenamefont{Mast, Dahm, and
  Fetter}}]{Mast85}
\bibinfo{author}{\bibfnamefont{D.~B.} \bibnamefont{Mast}},
  \bibinfo{author}{\bibfnamefont{A.~J.} \bibnamefont{Dahm}}, \bibnamefont{and}
  \bibinfo{author}{\bibfnamefont{A.~L.} \bibnamefont{Fetter}},
  \bibinfo{journal}{Phys. Rev. Lett.} \textbf{\bibinfo{volume}{54}},
  \bibinfo{pages}{1706} (\bibinfo{year}{1985}).

\bibitem[{\citenamefont{Glattli et~al.}(1985)\citenamefont{Glattli, Andrei,
  Deville, Poitrenaud, and Williams}}]{Glattli85}
\bibinfo{author}{\bibfnamefont{D.~C.} \bibnamefont{Glattli}},
  \bibinfo{author}{\bibfnamefont{E.~Y.} \bibnamefont{Andrei}},
  \bibinfo{author}{\bibfnamefont{G.}~\bibnamefont{Deville}},
  \bibinfo{author}{\bibfnamefont{J.}~\bibnamefont{Poitrenaud}},
  \bibnamefont{and} \bibinfo{author}{\bibfnamefont{F.~I.~B.}
  \bibnamefont{Williams}}, \bibinfo{journal}{Phys. Rev. Lett.}
  \textbf{\bibinfo{volume}{54}}, \bibinfo{pages}{1710} (\bibinfo{year}{1985}).

\bibitem[{\citenamefont{Volkov and Mikhailov}(1985)}]{Volkov85b}
\bibinfo{author}{\bibfnamefont{V.~A.} \bibnamefont{Volkov}} \bibnamefont{and}
  \bibinfo{author}{\bibfnamefont{S.~A.} \bibnamefont{Mikhailov}},
  \bibinfo{journal}{Pis'ma Zh. Eksp. Teor. Fiz.} \textbf{\bibinfo{volume}{42}},
  \bibinfo{pages}{450} (\bibinfo{year}{1985}), \bibinfo{note}{[JETP Lett. {\bf
  42}, 556-560 (1985)]}.

\bibitem[{\citenamefont{Volkov et~al.}(1986)\citenamefont{Volkov, Galchenkov,
  Galchenkov, Grodnenskii, Matov, and Mikhailov}}]{Volkov86b}
\bibinfo{author}{\bibfnamefont{V.~A.} \bibnamefont{Volkov}},
  \bibinfo{author}{\bibfnamefont{D.~V.} \bibnamefont{Galchenkov}},
  \bibinfo{author}{\bibfnamefont{L.~A.} \bibnamefont{Galchenkov}},
  \bibinfo{author}{\bibfnamefont{I.~M.} \bibnamefont{Grodnenskii}},
  \bibinfo{author}{\bibfnamefont{O.~R.} \bibnamefont{Matov}}, \bibnamefont{and}
  \bibinfo{author}{\bibfnamefont{S.~A.} \bibnamefont{Mikhailov}},
  \bibinfo{journal}{Pis'ma Zh. Eksp. Teor. Fiz.} \textbf{\bibinfo{volume}{44}},
  \bibinfo{pages}{510} (\bibinfo{year}{1986}), \bibinfo{note}{[JETP Lett. {\bf
  44}, 655-659 (1986)]}.

\bibitem[{\citenamefont{Fetter}(1985)}]{Fetter85}
\bibinfo{author}{\bibfnamefont{A.~L.} \bibnamefont{Fetter}},
  \bibinfo{journal}{Phys. Rev. B} \textbf{\bibinfo{volume}{32}},
  \bibinfo{pages}{7676} (\bibinfo{year}{1985}).

\bibitem[{\citenamefont{Fetter}(1986)}]{Fetter86a}
\bibinfo{author}{\bibfnamefont{A.~L.} \bibnamefont{Fetter}},
  \bibinfo{journal}{Phys. Rev. B} \textbf{\bibinfo{volume}{33}},
  \bibinfo{pages}{3717} (\bibinfo{year}{1986}).

\bibitem[{\citenamefont{Volkov and Mikhailov}(1988)}]{Volkov88}
\bibinfo{author}{\bibfnamefont{V.~A.} \bibnamefont{Volkov}} \bibnamefont{and}
  \bibinfo{author}{\bibfnamefont{S.~A.} \bibnamefont{Mikhailov}},
  \bibinfo{journal}{Zh. Eksp. Teor. Fiz.} \textbf{\bibinfo{volume}{94}},
  \bibinfo{pages}{217} (\bibinfo{year}{1988}), \bibinfo{note}{[Sov. Phys.-JETP
  {\bf 67}, 1639-1653 (1988)]}.

\bibitem[{\citenamefont{Allen et~al.}(1983)\citenamefont{Allen, St\"ormer, and
  Hwang}}]{Allen83}
\bibinfo{author}{\bibfnamefont{S.~J.} \bibnamefont{Allen}, \bibfnamefont{Jr.}},
  \bibinfo{author}{\bibfnamefont{H.~L.} \bibnamefont{St\"ormer}},
  \bibnamefont{and} \bibinfo{author}{\bibfnamefont{J.~C.~M.}
  \bibnamefont{Hwang}}, \bibinfo{journal}{Phys. Rev. B}
  \textbf{\bibinfo{volume}{28}}, \bibinfo{pages}{4875} (\bibinfo{year}{1983}).

\bibitem[{\citenamefont{Wassermeier et~al.}(1990)\citenamefont{Wassermeier,
  Oshinowo, Kotthaus, Macdonald, Foxon, and Harris}}]{Wassermeier90}
\bibinfo{author}{\bibfnamefont{M.}~\bibnamefont{Wassermeier}},
  \bibinfo{author}{\bibfnamefont{J.}~\bibnamefont{Oshinowo}},
  \bibinfo{author}{\bibfnamefont{J.~P.} \bibnamefont{Kotthaus}},
  \bibinfo{author}{\bibfnamefont{A.~H.} \bibnamefont{Macdonald}},
  \bibinfo{author}{\bibfnamefont{C.~T.} \bibnamefont{Foxon}}, \bibnamefont{and}
  \bibinfo{author}{\bibfnamefont{J.~J.} \bibnamefont{Harris}},
  \bibinfo{journal}{Phys. Rev. B} \textbf{\bibinfo{volume}{41}},
  \bibinfo{pages}{10287} (\bibinfo{year}{1990}).

\bibitem[{\citenamefont{Grodnensky and Kamaev}(1990)}]{Grodnensky90}
\bibinfo{author}{\bibfnamefont{I.~M.} \bibnamefont{Grodnensky}}
  \bibnamefont{and} \bibinfo{author}{\bibfnamefont{A.~Y.}
  \bibnamefont{Kamaev}}, \bibinfo{journal}{Surf. Sci.}
  \textbf{\bibinfo{volume}{229}}, \bibinfo{pages}{522} (\bibinfo{year}{1990}).

\bibitem[{\citenamefont{Volkov and Mikhailov}(1991)}]{Volkov91}
\bibinfo{author}{\bibfnamefont{V.~A.} \bibnamefont{Volkov}} \bibnamefont{and}
  \bibinfo{author}{\bibfnamefont{S.~A.} \bibnamefont{Mikhailov}}, in
  \emph{\bibinfo{booktitle}{Landau Level Spectroscopy}}, edited by
  \bibinfo{editor}{\bibfnamefont{G.}~\bibnamefont{Landwehr}} \bibnamefont{and}
  \bibinfo{editor}{\bibfnamefont{E.~I.} \bibnamefont{Rashba}}
  (\bibinfo{publisher}{North-Holland}, \bibinfo{address}{Amsterdam},
  \bibinfo{year}{1991}), vol. \bibinfo{volume}{27.2} of
  \emph{\bibinfo{series}{Modern Problems in Condensed Matter Sciences}},
  chap.~\bibinfo{chapter}{15}, pp. \bibinfo{pages}{855--907}.

\bibitem[{\citenamefont{Peters et~al.}(1991)\citenamefont{Peters, Lea, Janssen,
  Stone, Jacobs, Fozooni, and van~der Heijden}}]{Peters91}
\bibinfo{author}{\bibfnamefont{P.~J.~M.} \bibnamefont{Peters}},
  \bibinfo{author}{\bibfnamefont{M.~J.} \bibnamefont{Lea}},
  \bibinfo{author}{\bibfnamefont{A.~M.~L.} \bibnamefont{Janssen}},
  \bibinfo{author}{\bibfnamefont{A.~O.} \bibnamefont{Stone}},
  \bibinfo{author}{\bibfnamefont{W.~P. N.~M.} \bibnamefont{Jacobs}},
  \bibinfo{author}{\bibfnamefont{P.}~\bibnamefont{Fozooni}}, \bibnamefont{and}
  \bibinfo{author}{\bibfnamefont{R.~W.} \bibnamefont{van~der Heijden}},
  \bibinfo{journal}{Phys. Rev. Lett.} \textbf{\bibinfo{volume}{67}},
  \bibinfo{pages}{2199} (\bibinfo{year}{1991}).

\bibitem[{\citenamefont{Ashoori et~al.}(1992)\citenamefont{Ashoori, St\"ormer,
  Pfeiffer, Baldwin, and West}}]{Ashoori92}
\bibinfo{author}{\bibfnamefont{R.~C.} \bibnamefont{Ashoori}},
  \bibinfo{author}{\bibfnamefont{H.~L.} \bibnamefont{St\"ormer}},
  \bibinfo{author}{\bibfnamefont{L.~N.} \bibnamefont{Pfeiffer}},
  \bibinfo{author}{\bibfnamefont{K.~W.} \bibnamefont{Baldwin}},
  \bibnamefont{and} \bibinfo{author}{\bibfnamefont{K.}~\bibnamefont{West}},
  \bibinfo{journal}{Phys. Rev. B} \textbf{\bibinfo{volume}{45}},
  \bibinfo{pages}{3894} (\bibinfo{year}{1992}).

\bibitem[{\citenamefont{Mikhailov and Volkov}(1992)}]{Mikhailov92}
\bibinfo{author}{\bibfnamefont{S.~A.} \bibnamefont{Mikhailov}}
  \bibnamefont{and} \bibinfo{author}{\bibfnamefont{V.~A.}
  \bibnamefont{Volkov}}, \bibinfo{journal}{J. Phys. Condens. Matter}
  \textbf{\bibinfo{volume}{4}}, \bibinfo{pages}{6523} (\bibinfo{year}{1992}).

\bibitem[{\citenamefont{Talyanskii et~al.}(1993)\citenamefont{Talyanskii,
  Frost, Pepper, Ritchie, Grimshaw, and Jones}}]{Talyanskii93}
\bibinfo{author}{\bibfnamefont{V.~K.} \bibnamefont{Talyanskii}},
  \bibinfo{author}{\bibfnamefont{J.~E.~F.} \bibnamefont{Frost}},
  \bibinfo{author}{\bibfnamefont{M.}~\bibnamefont{Pepper}},
  \bibinfo{author}{\bibfnamefont{D.~A.} \bibnamefont{Ritchie}},
  \bibinfo{author}{\bibfnamefont{M.}~\bibnamefont{Grimshaw}}, \bibnamefont{and}
  \bibinfo{author}{\bibfnamefont{G.~A.~C.} \bibnamefont{Jones}},
  \bibinfo{journal}{J. Phys. Condens. Matter} \textbf{\bibinfo{volume}{5}},
  \bibinfo{pages}{7643} (\bibinfo{year}{1993}).

\bibitem[{\citenamefont{Zhitenev et~al.}(1993)\citenamefont{Zhitenev, Haug, von
  Klitzing, and Eberl}}]{Zhitenev93}
\bibinfo{author}{\bibfnamefont{N.~B.} \bibnamefont{Zhitenev}},
  \bibinfo{author}{\bibfnamefont{R.~J.} \bibnamefont{Haug}},
  \bibinfo{author}{\bibfnamefont{K.}~\bibnamefont{von Klitzing}},
  \bibnamefont{and} \bibinfo{author}{\bibfnamefont{K.}~\bibnamefont{Eberl}},
  \bibinfo{journal}{Phys. Rev. Lett.} \textbf{\bibinfo{volume}{71}},
  \bibinfo{pages}{2292} (\bibinfo{year}{1993}).

\bibitem[{\citenamefont{Aleiner and Glazman}(1994)}]{Aleiner94}
\bibinfo{author}{\bibfnamefont{I.~L.} \bibnamefont{Aleiner}} \bibnamefont{and}
  \bibinfo{author}{\bibfnamefont{L.~I.} \bibnamefont{Glazman}},
  \bibinfo{journal}{Phys. Rev. Lett.} \textbf{\bibinfo{volume}{72}},
  \bibinfo{pages}{2935} (\bibinfo{year}{1994}).

\bibitem[{\citenamefont{Tonouchi et~al.}(1994)\citenamefont{Tonouchi, Miyasato,
  Hawker, Cheng, and Rampton}}]{Tonouchi94}
\bibinfo{author}{\bibfnamefont{M.}~\bibnamefont{Tonouchi}},
  \bibinfo{author}{\bibfnamefont{T.}~\bibnamefont{Miyasato}},
  \bibinfo{author}{\bibfnamefont{P.}~\bibnamefont{Hawker}},
  \bibinfo{author}{\bibfnamefont{T.~S.} \bibnamefont{Cheng}}, \bibnamefont{and}
  \bibinfo{author}{\bibfnamefont{V.~W.} \bibnamefont{Rampton}},
  \bibinfo{journal}{J. Phys. Soc. Japan} \textbf{\bibinfo{volume}{63}},
  \bibinfo{pages}{4499} (\bibinfo{year}{1994}).

\bibitem[{\citenamefont{Dahl et~al.}(1995)\citenamefont{Dahl, Manus, Kotthaus,
  Nickel, and Schlapp}}]{Dahl95}
\bibinfo{author}{\bibfnamefont{C.}~\bibnamefont{Dahl}},
  \bibinfo{author}{\bibfnamefont{S.}~\bibnamefont{Manus}},
  \bibinfo{author}{\bibfnamefont{J.~P.} \bibnamefont{Kotthaus}},
  \bibinfo{author}{\bibfnamefont{H.}~\bibnamefont{Nickel}}, \bibnamefont{and}
  \bibinfo{author}{\bibfnamefont{W.}~\bibnamefont{Schlapp}},
  \bibinfo{journal}{Appl. Phys. Lett.} \textbf{\bibinfo{volume}{66}},
  \bibinfo{pages}{2271} (\bibinfo{year}{1995}).

\bibitem[{\citenamefont{Mikhailov}(1995)}]{Mikhailov95a}
\bibinfo{author}{\bibfnamefont{S.~A.} \bibnamefont{Mikhailov}},
  \bibinfo{journal}{Pis'ma Zh. Eksp. Teor. Fiz.} \textbf{\bibinfo{volume}{61}},
  \bibinfo{pages}{412} (\bibinfo{year}{1995}), \bibinfo{note}{[JETP Lett. {\bf
  61}, 418-423 (1995)]}.

\bibitem[{\citenamefont{Balaban et~al.}(1998)\citenamefont{Balaban, Meirav, and
  Bar-Joseph}}]{Balaban98}
\bibinfo{author}{\bibfnamefont{N.~Q.} \bibnamefont{Balaban}},
  \bibinfo{author}{\bibfnamefont{U.}~\bibnamefont{Meirav}}, \bibnamefont{and}
  \bibinfo{author}{\bibfnamefont{I.}~\bibnamefont{Bar-Joseph}},
  \bibinfo{journal}{Phys. Rev. Lett.} \textbf{\bibinfo{volume}{81}},
  \bibinfo{pages}{4967} (\bibinfo{year}{1998}).

\bibitem[{\citenamefont{Mikhailov}(2000)}]{Mikhailov00a}
\bibinfo{author}{\bibfnamefont{S.~A.} \bibnamefont{Mikhailov}}, in
  \emph{\bibinfo{booktitle}{Edge Excitations of Low-Dimensional Charged
  Systems}}, edited by
  \bibinfo{editor}{\bibfnamefont{O.}~\bibnamefont{Kirichek}}
  (\bibinfo{publisher}{Nova Science Publishers, Inc.}, \bibinfo{address}{NY},
  \bibinfo{year}{2000}), chap.~\bibinfo{chapter}{1}, pp.
  \bibinfo{pages}{1--47}.

\bibitem[{\citenamefont{Dorozhkin et~al.}(2005)\citenamefont{Dorozhkin,
  Tovstonog, Mikhailov, Kukushkin, Smet, and von Klitzing}}]{Dorozhkin05}
\bibinfo{author}{\bibfnamefont{P.~S.} \bibnamefont{Dorozhkin}},
  \bibinfo{author}{\bibfnamefont{S.~V.} \bibnamefont{Tovstonog}},
  \bibinfo{author}{\bibfnamefont{S.~A.} \bibnamefont{Mikhailov}},
  \bibinfo{author}{\bibfnamefont{I.~V.} \bibnamefont{Kukushkin}},
  \bibinfo{author}{\bibfnamefont{J.~H.} \bibnamefont{Smet}}, \bibnamefont{and}
  \bibinfo{author}{\bibfnamefont{K.}~\bibnamefont{von Klitzing}},
  \bibinfo{journal}{Appl. Phys. Lett.} \textbf{\bibinfo{volume}{87}},
  \bibinfo{pages}{092107} (\bibinfo{year}{2005}).

\bibitem[{\citenamefont{Chiu and Quinn}(1974)}]{Chiu74}
\bibinfo{author}{\bibfnamefont{K.~W.} \bibnamefont{Chiu}} \bibnamefont{and}
  \bibinfo{author}{\bibfnamefont{J.~J.} \bibnamefont{Quinn}},
  \bibinfo{journal}{Phys. Rev. B} \textbf{\bibinfo{volume}{9}},
  \bibinfo{pages}{4724} (\bibinfo{year}{1974}).

\end{thebibliography}


\end{document}